\documentclass[11pt,a4wide]{article}

\usepackage{hyperref}
\usepackage{amsmath}
\usepackage{amsfonts}
\usepackage{amsbsy}
\usepackage{epsfig}
\usepackage{latexsym}
\usepackage{xcolor}
\input amssym.def
\input amssym.tex
\usepackage{graphicx}

\advance \topmargin by -\headheight
\advance \topmargin by -\headsep
\evensidemargin \oddsidemargin
\advance\hoffset by -3mm  
\advance\voffset by  8mm  
\def\bbbc{{\mathchoice {\setbox0=\hbox{$\displaystyle\rm C$}\hbox{\hbox
to0pt{\kern0.4\wd0\vrule height0.9\ht0\hss}\box0}}
{\setbox0=\hbox{$\textstyle\rm C$}\hbox{\hbox
to0pt{\kern0.4\wd0\vrule height0.9\ht0\hss}\box0}}
{\setbox0=\hbox{$\scriptstyle\rm C$}\hbox{\hbox
to0pt{\kern0.4\wd0\vrule height0.9\ht0\hss}\box0}}
{\setbox0=\hbox{$\scriptscriptstyle\rm C$}\hbox{\hbox
to0pt{\kern0.4\wd0\vrule height0.9\ht0\hss}\box0}}}}

\makeatletter
\@addtoreset{equation}{section}
\makeatother

\begin{document}
\title{{\Large \bf{Algorithms for Generating All Static Spherically Symmetric (An)isotropic Fluid Solutions of Einstein's Equations}}}
\author{
M M Akbar$^{}$\footnote{E-mail: akbar@utdallas.edu}
\ \,\,\&\,
R Solanki$^{}$\footnote{E-mail: rahulkumar.solanki@utdallas.edu}
\\
\\ {$^{*}${Department of Mathematical Sciences}}
\\$^{\dagger}${Department of Physics}
\\ {University of Texas at Dallas}
\\ {Richardson, TX 75080, USA}}
\date{\today}
 \maketitle
\begin{abstract}
\noindent
We study the Einstein equations of the static spherically symmetric anisotropic fluid system in curvature coordinates to find algorithms that generate all solutions and all solutions that are regular at the center. All possible combinations of input functions from the set of four functions that characterize the anisotropic system are considered and all equivalent conditions for central regularity are determined (for both isotropic and anisotropic systems). We provide the first regularity analysis of the known algorithm that uses the potential function and anisotropy as inputs. The one-parameter family of solutions produced by this algorithm and by its isotropic counterpart are regular at the center for the usual initial conditions on the input functions; these conditions by themselves are inadequate for geometric regularity, but the Einstein equations fill the gap. By slightly reformulating the anisotropic algorithm, all regular solutions can be obtained in a 1--1 relation with the input.  This allows us to interpret the parameter for the isotropic and anisotropic systems as the central density. For three other choices of input function pairs (any two of the potential function, density, or radial pressure), a remarkably straightforward algorithm follows, which is very efficient in generating regular anisotropic solutions. This is because the equivalency of the three pairs in this algorithm arises precisely from the same algebraic relation that made the different equivalent sets of regularity conditions possible. In addition, the choice of functions makes this algorithm very suitable for finding particular solutions that admit other desirable physical properties; we construct three examples. This algorithm does not admit an isotropic limit although all isotropic solutions are produced as part of the anisotropic system. The remaining two choices of input function pairs (anisotropy with the radial pressure or density) lead to the old barriers one encounters in the isotropic system: Riccati and Abel equations. However, with any solution generated by the new and existing algorithms, one can now construct the general solution of the corresponding Riccati equation to obtain a one-parameter family of geometries for each input solution. We discuss the regularity of the resulting solutions.
\end{abstract}

\section{Introduction}\label{sec:i}

The most general spherically symmetric static-fluid configurations possible in general relativity have anisotropic pressure. Einstein's equations for this system reduce to a system of two equations in four unknown functions: density, radial pressure, (one) tangential pressure, and the potential function. The difference between the two pressures is the anisotropy function, and when this vanishes identically, one obtains the system of isotropic fluid configurations. Isotropic and anisotropic systems, thus, require one and two input functions, respectively, to produce a solution. One can also use an equation of state as an input. Isotropic fluid spheres (or perfect fluid spheres, as they are more commonly known) have been studied extensively since the early days of relativity and continue to be of physical and mathematical interest \cite{kramer1980exact,finch1989painleve,delgaty1998physical}. There is a large corpus of exact solutions of the isotropic system, which were obtained by employing various ad hoc techniques. However, only a very few of them can be considered physical \cite{delgaty1998physical} and, thus, to obtain physically reasonable solutions, one generally uses numerical integration. Despite Einstein's cluster solution in the late 30s \cite{einstein1939stationary}, anisotropic fluid spheres have received much less attention. In recent times, they have attracted fresh interest, as anisotropy has been found to be relevant in several physical situations \cite{Herrera:1997plx}, especially in modeling static galactic dark matter halos (see, for example, \cite{boehmer2007einstein}) as well as compact stars, both of which are currently attracting a considerable amount of research interest \cite{bohmer2006bounds,dev2002anisotropic,viaggiu2009modeling}.

More recently, a number of generation algorithms have been developed that can obtain all static isotropic fluid spheres by quadrature from a single function---the generating function---in various coordinates \cite{berger1987general, PhysRevD.67.104015, Rahman:2001hp, Martin:2003jc}. This approach is comparable to the well-studied static axisymmetric vacuum system for which one can generate all solutions from an arbitrary (harmonic) generating function (see, for example, \cite{kramer1980exact}). An algorithm gives considerable insight into the structure of the solution space and can be useful in constructing new solutions, exact or otherwise, of particular interest. There is also an algorithm for the anisotropic system in canonical coordinates \cite{herrera2008all}, which extends the perfect fluid algorithm in \cite{PhysRevD.67.104015}. The two inputs used in this algorithm are the potential function (equivalently, the $g_{tt}$ component of the metric) and the anisotropy. Another algorithm uses two inputs that are functions of fluid variables to generate anisotropic solutions in canonical coordinates. A less general algorithm that generates a class of anisotropic solutions from isotropic Newtonian solutions is known \cite{lake2009generating}. Generating functions are not \emph{a priori} required to be quantities that are readily physically interpretable and they could be complicated expressions in terms of the physical quantities of the system (see, for example, \cite{Fodor:2000gu}). In any case, even when they can produce all solutions, it is not easy to determine beforehand what inputs will give a desired output solution (like solutions that are regular at the center and have other properties of physical interest). Addressing such issues will be central to this paper.

In this paper, we will systematically study all possible generation algorithms in canonical coordinates using the basic functions of the (an)isotropic system (as in the algorithms of \cite{PhysRevD.67.104015} and \cite{herrera2008all}). We will not consider possibilities using other functions as inputs (as in \cite{Fodor:2000gu, Martin:2003jc}). A generating algorithm that finds all solutions, or solutions of  a certain type, of the system via quadrature is not in general not the best way to find \emph{particular} solutions and these are generally considered separate questions. However, using the basic variables of the system, they can be simultaneously approached since physical conditions in canonical coordinates are given in terms of the basic variables and their derivatives. As we will see in this paper, regularity can be considered within a new algorithm which generates all regular solutions in canonical coordinates. This will streamline the search for particular solutions that satisfy other physical properties. We have three main goals: (i) to identify \emph{all} possible algorithms using these basic variables such that each algorithm generates all solutions of the system via quadrature, (ii) to generate all (and only) regular solutions using each algorithm found via (i),\footnote{Note that generating solutions that are regular at the center (i.e., scalar polynomial invariants constructed from that Riemann tensor that are finite) is an altogether different problem from checking whether an existing solution is regular. There are no discussions in the literature on the regularity of the algorithm in \cite{herrera2008all}. We will address this as well as regularity in all other algorithms we find. } and (iii) to use (i) and (ii) to find particular examples that satisfy other physical conditions.


This paper is arranged as follows. In Section~\ref{sec:ii}, we introduce the Einstein equations for the system, and in Section~\ref{sec:iii} we examine the geometric conditions of regularity in the light of Einstein's equations. In Section~\ref{sec:iv}, we study the existing algorithm(s) and reformulate it in terms of the tangential pressure function and show how both can produce solutions that are regular at the center. We then present a new algorithm that generates all solutions from three other pairs of functions; this algorithm makes generating regular anisotropic solutions very straightforward. Then, in Section~\ref{sec:v}, we reconsider the Riccati equations---which on their own could not be integrated---in light of the algorithms. In Section~\ref{sec:vi}, we find solutions that satisfy other physical conditions, staring from the new algorithm. We conclude in Section~\ref{sec:vii}.

\section{Spherically Symmetric Static Fluid Systems} \label{sec:ii}

We will be working in the curvature/Schwarzschild coordinates \cite{misner2017gravitation}:
\begin{equation}
\label{line_element0}
ds^2= -e^{2\Phi(r)} dt^2+ e^{2\Psi(r)} dr^2 +r^2 (d\theta^2+\sin^2{\theta} d\phi^2),
\end{equation}
for which the most general fluid energy--momentum tensor is (see, for example, \cite{ellis1971relativistic})
\begin{equation}
\label{EM_imperfect}
T^{t}_{t}=-\rho(r), \quad T^{r}_{r}=p(r), \quad T^{\theta}_{\theta}=T^{\phi}_{\phi}=P(r),
\end{equation}
where $\rho(r)$ is the energy density, $p(r)$ is the radial pressure, and $P(r)$ is the tangential pressure. The anisotropy is then measured by the anisotropic pressure function: $\chi(r)\equiv {(P(r)-p(r))}/{3}$. This is often used instead of $P(r)$. Integrating $G^{t}_{t}=8\pi T^{t}_{t}$ from the origin, we get the ``mass" function:
\begin{equation}
\label{density}
m(r)=\int^{r}_0 4\pi \rho(x)x^2 dx, \quad \frac{dm}{dr}=4\pi \rho(r)r^2.
\end{equation}
One can also consider a ``core" for $r \le r_c\ne0$ (which could be made of another fluid solution, a vacuum Schwarzschild solution, or something else):
\begin{equation}
\label{density}
m(r)=\int^{r}_{r_c} 4\pi \rho(x)x^2 dx + m_c, \quad \frac{dm}{dr}=4\pi \rho(r)r^2,
\end{equation}
where $m_c$ is the ``core mass." In either case, we get the following ``simplified" line element:
\begin{equation}
\label{line_element}
ds^2= -e^{2\Phi(r)} dt^2+ \frac{dr^2}{1-2m(r)/r} +r^2 (d\theta^2+\sin^2{\theta} d\phi^2).
\end{equation}
For the remaining two Einstein equations, it is customary to use the conservation equation, $\nabla_{\mu}T^{\mu}_{r}=0$, and $G^{r}_{r}=8\pi T^{r}_{r}$ to obtain
\begin{equation}
\label{momentum_conservation}
P=\frac{r}{2}\left(\frac{dp}{dr}+(\rho+p)\frac{d\Phi}{dr}\right)+p
\end{equation}
and
\begin{equation}
\label{EFE}
\frac{d\Phi}{dr}-\frac{m+4\pi r^3 p}{r(r-2m)}=0,
\end{equation}
respectively. The algebraic implication of this equation for the anisotropic system will become important later in this paper.

Substituting~(\ref{EFE}) into~(\ref{momentum_conservation}) gives the generalized Tolman--Oppenheimer--Volkoff (TOV) equation:
\begin{equation}
\label{generalized_TOV1}
\frac{dp}{dr}+\left[{m+4\pi\rho r^3 \over r(r-2m)} + \frac{2}{r}\right]p+\left(\frac{4\pi r^2}{r-2m}\right)p^2=\frac{2P}{r}-\frac{\rho m}{r(r-2m)},
\end{equation}
which takes a slightly tidier form with $\chi(r)$:
\begin{equation}
\label{generalized_TOV11}
\frac{dp}{dr}+\left[{m+4\pi\rho r^3 \over r(r-2m)}\right]p+\left(\frac{4\pi r^2}{r-2m}\right)p^2=\frac{6\chi}{r}-\frac{\rho m}{r(r-2m)}.
\end{equation}
In the following, we will often refer to (\ref{EFE}) and (\ref{generalized_TOV1}) as ``the system" and will switch between (\ref{generalized_TOV1}) and (\ref{generalized_TOV11}) freely. The perfect or isotropic system is obtained by setting  $p(r)=P(r)$ or $\chi(r)=0$ in the generalized TOV equation and will be referred to as the ``isotropic system.''
\section{Central Regularity}\label{sec:iii}
The simplified line element (\ref{line_element}) has reduced the number of independent ordinary differential equations to two, equations (\ref{momentum_conservation}) and (\ref{EFE}) or equations (\ref{EFE}) and (\ref{generalized_TOV1}), in four unknowns, and thus two inputs are needed to integrate the system (one for the isotropic case). In stellar models with isotropic fluids, for example, the equation of state $p=p(\rho)$ acts as the functional input needed. One takes a finite initial value $\rho(0)$  and integrates (\ref{density}) and (\ref{generalized_TOV11}) (with $\chi=0$) simultaneously until a radius $r=r_b$ is reached where $p(r)$ vanishes (see, for example, \cite{Hartlebook, misner2017gravitation}).\footnote{One also generally requires the solution to be surrounded by a Schwarzschild exterior solution (i.e., $T^{\mu}_{\nu}=0\;\text{for}\;r>r_b$). Continuity of the metric components requires that $e^{2\Phi(r_b)}=\left(1-{2M}/{r_b}\right)$, where $M=m(r_b)$ is the mass of the Schwarzschild solution. Moreover, the differentiability of the metric components requires the continuity of $p(r)$ and $\rho(r)$. However, physically, for hydrostatic equilibrium, the radial pressure is required to be continuous across the boundary, i.e., $p(r_b)=0$, but $\rho(r_b)$ can be nonzero (see e.g., \cite{Bowers:1974tgi}).}
Any initial finite value of $\rho(0)$, which returns a finite central pressure via the equation of state $p(0)=p(\rho(0))$, ensures regularity at the center.

We will be interested in generating all solutions via solution-generating algorithms --- infinite fluid spheres as well as those that terminate at finite radii, as in stellar models. We will also pay special attention to generating solutions that are regular at the center. As we will discuss in detail later, in any solution generation algorithm the generating functions are the input functions which are freely specifiable and which determine the remaining functions via quadratures. For the resulting solutions to be regular at the center one has to ensure that: (i) the input/generating functions have all the properties consistent with the geometric condition of regularity and (ii) their initial values also ensure that the initial values of the remaining variables are consistent with regularity. If these conditions cannot be fulfilled one will not be able to specialize to regular solutions, although one may be able to generate \emph{all} solutions. With these in mind, we take a closer look at the regularity conditions below.

As usual in the literature, by regularity, we mean the finiteness of scalar polynomial invariants constructed from the Riemann tensor. This has been discussed by a number of authors for both isotropic and anisotropic fluid spheres (see, for example, \cite{Bronnikov:2012wsj, lake_regularity, delgaty1998physical,lake2009generating}). For spherically symmetric static spacetimes it is sufficient that the Kretschmann scalar $\kappa=R_{\mu\nu}^{\;\;\;\;\rho\sigma}R_{\rho\sigma}^{\;\;\;\;\mu\nu}$ is finite for scalar
polynomial invariants constructed from the Riemann tensor (see, for example, \cite{Bronnikov:2012wsj}).

For such a spacetime in static coordinates, like (\ref{line_element0}) and (\ref{line_element}), the algebraically independent Riemann tensor is pairwise diagonal with the following nonzero components: $R_{01}^{\;\;\;01}, R_{02}^{\;\;\;02}=R_{03}^{\;\;\;03}, R_{12}^{\;\;\;12}=R_{13}^{\;\;\;13}, R_{23}^{\;\;\;23}$.\footnote{For a line element in stationary coordinates where $g_{01}\ne 0$,  for example, $R_{12}^{\;\;\;02} \neq 0$.}
Thus, the Kretschmann scalar becomes a sum of four squares:
\begin{equation}
\kappa=4K_{1}^2+8K_{2}^2+8K_{3}^2+4K_{4}^2,
\end{equation}
where $K_1=-R_{01}^{\;\;\;01}$, $K_2=-R_{02}^{\;\;\;02}$, $K_3=-R_{13}^{\;\;\;13}$ and $K_4=-R_{23}^{\;\;\;23}$. This is finite iff each square term is finite.\footnote{Recall, for the Schwarzschild exterior, that $R_{1212}=-M/(r-2M)$ i.e., singular at $r=2M$ but $R_{12}^{\;\;\;12}=-M/r^3$.}.
Thus, the finiteness of $\kappa$ is sufficient to ensure the regularity of these scalars \cite{lake_regularity}.

In the simplified line element (\ref{line_element}), one geometric and one fluid variables appear, $\Phi(r)$ and $\rho(r)$, and\footnote{The tensor calculations were performed with Maple using the GRTensor III package.}
\begin{multline}
\kappa = 4\left[\left(1-{2m \over r}\right)(\Phi''+\Phi'^2)+\Phi'\left(\frac{m}{r^2}-4\pi\rho r\right)\right]^2 \\
  + 8\left[\left(1-\frac{2m}{r}\right)\frac{\Phi'}{r}\right]^2
  + 8\left({m \over r^3}-4\pi\rho\right)^2+4\left(\frac{2m}{r^3}\right)^2.
\end{multline}
As $r\rightarrow 0$, ${m}/{r^3} \rightarrow 4\pi \rho/3$ from the definition of $m(r)$, equation (\ref{density}). The last two terms in $\kappa$ near $r=0$ are separately proportional to $\rho$, so one needs $\rho(0)$ to be finite. For the second term to be finite at the center, one needs $\Phi'(0)=0$ and $\Phi''(0)$ finite simultaneously, which make the first term finite without any extra conditions. Thus, the necessary and sufficient conditions that $\kappa$ is finite at $r=0$ are
\begin{eqnarray}
\label{regulargeomtric}
&& \rho(0) \text{ finite},\\
&& \Phi'(0)=0 \text{ and } \Phi''(0) \text{ finite}.
\end{eqnarray}
Under the above conditions, the first and second Weyl invariants, given by $(C(r))^2/6$ and $(C(r))^3/36$, respectively, where
\begin{eqnarray}
	 C(r)=\left(1-\frac{2m}{r}\right)(\Phi''+\Phi'^2)+4\pi\rho-\frac{3m}{r^3}-\frac{\Phi'}{r}\left(4\pi\rho r^2-{3m \over r}+1\right), \label{C_geometry}
	\end{eqnarray}
vanish identically as $r\rightarrow 0$. This shows that the metric is conformally flat at the origin under these conditions \cite{delgaty1998physical, lake_regularity}. These conditions are geometric in that none of the Einstein equations except the $G^t_t$ component has been used, which amounts to merely renaming $\Psi(r)$. These apply to both isotropic and anisotropic fluids spheres.
\subsubsection*{Equivalent Conditions of Regularity}
We now explore the consequence of these geometric conditions on the other variables of the system. Applying them to the $G^r_r$ and $G^\theta_\theta$/$G^\phi_\phi$ equations,\footnote{\begin{equation}
\label{GrrEinstein}
G^{r}_{r}\equiv 2 \left(1-\frac{2m}{r}\right) \frac{\Phi'(r)}{r}-\frac{2m}{r^3}=8\pi p(r)
\end{equation}
and
\begin{equation}
\label{EEthetatheta}
G^{\theta}_{\theta}=G^{\phi}_{\phi}=\left(1-\frac{2m}{r}\right)\left(\Phi''+\frac{\Phi'}{r}+\Phi'^2\right)+\left(1+r\Phi'\right)\left(\frac{m}{r^3}-4\pi\rho\right)=8\pi P(r).
\end{equation}}
we find
\begin{eqnarray}
\label{algebraicrelation}
&& p(0)=\frac{1}{4\pi}\Phi''(0)-\frac{1}{3}\rho(0),\\\label{algebraicrelationP}
&& P(0)=\frac{1}{4\pi}\Phi''(0)-\frac{1}{3}\rho(0).
\end{eqnarray}
The first relation between nonzero finite quantities is known \cite{PhysRevD.67.104015} and they will be useful for our analysis. In terms of anisotropy one can write them as
\begin{eqnarray}
\label{algebraicrelation1}
&& p(0)=\frac{1}{4\pi}\Phi''(0)-\frac{1}{3}\rho(0),\\
&& \chi(0)=0.
\end{eqnarray}
Note that any two of $\rho(0)$, $p(0)$, and $\Phi''(0)$ will fix the initial value of the third quantity via (\ref{algebraicrelation}) as well as imply that $\Phi'(0)=0$ and $P(0)=p(0)$ (or $\chi(0)=0$) as a consequence of the Einstein equations. In the isotropic case, $P(0)=p(0)$ holds trivially and finiteness of $\rho(0)$ ensures finiteness of $p(0)$ via an equation of state as in stellar models. Thus, the following combinations are equivalent for regularity:
\begin{eqnarray}
\label{correctinpus0}
&&\rho(0)\,\,\,\mathrm{finite}, \,\,\, \Phi'(0)=0,\,\,\,\Phi''(0)\,\,\,\mathrm{finite}\\\label{correctinpus1}
&&p(0)\,\,\,\mathrm{finite}, \,\,\, \Phi'(0)=0, \,\,\,\Phi''(0)\,\,\, \mathrm{finite} \\\label{correctinpus2}
&& \rho(0)\,\,\,\mathrm{finite}, \,\,\, p(0)\,\,\,\mathrm{finite}
\end{eqnarray}
Note that none of the above uses $P(0)$. Combinations of $P(0)$ with other functions like
\begin{eqnarray}
\label{correctinpus3}
&&P(0)\,\,\,\mathrm{finite}, \,\,\, \Phi'(0)=0,\,\,\, \Phi''(0)\,\,\, \mathrm{finite} \\
\label{correctinpus4}
&&\rho(0)\,\,\,\mathrm{finite}, \,\,\, P(0)\,\,\, \mathrm{finite}
\end{eqnarray}
are not sufficient to reproduce the geometric conditions of regularity. One, thus, needs a supplemental condition to be able to use them. As we will see when generating regular solutions using $(\Phi'(r), P(r))$, such a condition arises quite fortuitously from the considerations of the differential equation. We will also see that
\begin{eqnarray}
\label{correctinpus5}
&&\chi(0)=0,\,\,\, \Phi'(0)=0,\,\,\, \Phi''(0)\,\,\, \mathrm{finite}
\end{eqnarray}
and
\begin{eqnarray}
\label{correctinpus6}
&&\Phi'(0)=0,\,\,\, \Phi''(0)\,\,\, \mathrm{finite}
\end{eqnarray}
can serve as sufficient conditions for regularity for the anisotropic and isotropic systems, respectively. This is because Einstein's equations will produce the right behavior for $m(r)$, which ensures that $\kappa$ is finite at the origin.
\subsubsection*{TOV and Momentum Conservation with Regularity}
The only quantity that has not appeared so far in the discussion is $p'(r)$. To see the effect of regularity on $p'(r)$, it is natural to turn to the generalized TOV equation.\footnote{The TOV equation can be turned into a differential equation in  $m(r)$ and $\Phi(r)$, as we will see later.} For the anisotropic case, it is easy to see that the TOV applies only for $r >0$. However, we argue that, even for the isotropic case, one needs to look at the original momentum conservation equation (\ref{momentum_conservation}) where $p'(r)$ appears only in $rp'(r)$, which is clearly valid at $r=0$. The conditions of regularity then imply that $p'(r)$ will satisfy the following condition:
\begin{equation}
\label{derivative}
\lim_{r\to 0} r p'(r)=0 \,\,\,\,(\text{isotropy and anisotropy})
\end{equation}
under the above regularity conditions. This condition can be fulfilled by any finite $p'(0)$ (including zero), which would have been the only conclusion had we used the TOV equation. However, (\ref{derivative}) clearly shows $p(r)$ does not have to be differentiable at $r=0$ while $p(0)$ is finite. Such possibilities should be included if one is interested in generating all regular solutions.

Further corroboration of condition (\ref{derivative}) comes from casting the Kretschmann scalar in terms of $p(r)$ using (\ref{EFE}):
\begin{multline}
	\kappa = 4\left(4\pi (\rho+p) -\frac{2m}{r^3} + rp'(r) + \frac{(m+4\pi r^3 p)(p+\rho)}{r-2m} \right)^2 \\
   + 8\left({m \over r^3}+4\pi p\right)^2
   + 8\left({m \over r^3}-4\pi\rho\right)^2
   + 4\left(\frac{2m}{r^3}\right)^2.
\end{multline}
The last two terms are the same as before and the second term implies that $p(0)$ should be finite and that $\lim_{r\to 0} r p'(r)$ is \emph{finite}. However, for conformal flatness at the center, the vanishing of
\begin{eqnarray}
	 C(r)= 4\pi \left(rp'(r)+\frac{(m+4\pi r^3 p)(p+\rho)}{r-2m}+  2\rho-\frac{6m}{4\pi r^3} \right) \label{C_fluid}
\end{eqnarray}
requires $\lim_{r\to 0} r p'(r)=0$. Also note that, in either of the expressions above, $p'(r)$ does not appear except in $rp'(r)$. That the finiteness of $\kappa$ in terms of $\rho(r)$ and $p(r)$ does not guarantee conformal flatness at the center, unlike when $\rho(r)$ and $\Phi'(r)$ were used, shows that something can get lost in translation.

To summarize, in choosing $p(r)$ as a generating function, it has to be chosen such that
\begin{eqnarray}
\label{correctinpus}
p(0)\,\,\,\mathrm{finite}, \,\,\, \lim_{r\to 0} r p'(r)=0
\end{eqnarray}
to ensure all possible regular solutions. Other variables in (\ref{correctinpus0})--(\ref{correctinpus4}) do not need any such qualification except that they are continuous and differentiable as shown. It is easy to see that, given the three quantities in the algebraic equation (\ref{algebraicrelation}), no one function alone can ensure regularity at the center. This makes regularity in the anisotropic systems, where two inputs are needed, much easier to handle.
\\
\\
\emph{A Simple Nontrivial Example:} We have discussed above how it is natural to conclude that $p'(0)$ is finite if one uses the TOV equation. This narrative is reinforced by the fact that $p'(0)=0$ is often preferred physically. However, it is easy to see that $\lim_{r\to 0} r p'(r)=0$ (with finite $p(0)$) can be satisfied by functions like the following:
\begin{equation}
\label{exampleofp}
p(r)=c-r^{1/n} e^r,
\end{equation}
where $n>1$ and $c>0$ is a constant. It is monotonically decreasing and $\lim_{r\to 0^{+}} p'(r)= \mathrm{DNE}$, but $\lim_{r\to 0^{+}} r p'(r)= 0$ and $p(0)=c$. It is easy to construct similar examples. Such solutions should be included in the consideration of all regular solutions for both the isotropic and anisotropic systems.

\section{Generating Functions and Solutions}\label{sec:iv}

Although for any pair of input functions one can find a solution of the system (\ref{EFE}) and (\ref{generalized_TOV1}), for a pair of generating functions the system admits a general form solution and the metric can be expressed formally in terms of integrals involving these functions only. One then has what is called a generation algorithm.\footnote{Whether those integrations can be performed exactly to obtain an exact solution is considered an \emph{a posteriori} issue and considered individually. In most cases, this will require numerical integration, as mentioned in the introduction.}

As mentioned in the introduction, generating functions do not have to be from the set of basic variables, although this is the case in the existing algorithms in canonical coordinates, which we will revisit below \cite{PhysRevD.67.104015, herrera2008all}.
We will consider all six combinations of two inputs that are possible out of $\rho(r)$, $p(r)$, $P(r)$, and $\Phi(r)$. First, note that the following sets are equivalent:
\begin{enumerate}
  \item Linear combinations of $\rho(r)$, $p(r)$, and $P(r)$ are equivalent. This just involves rewriting the Einstein equations using corresponding linear combinations. This makes $P(r)$ and $\chi(r)$ equivalent inputs when used alongside $p(r)$, which we have seen in (\ref{generalized_TOV1}) and (\ref{generalized_TOV11}).\footnote{If $p(r)$ is not used, one may be preferred over the other, as we have seen in the regularity conditions (\ref{correctinpus4}).} However, if one wants to take advantage of the simplified line element (\ref{line_element}), one must keep $\rho(r)$ as it is, as we do.
  \item Specifying a function is equivalent to specifying another function if they are in a 1-1 algebraic correspondence. This obvious fact will help identify equivalent functions with the basic functions of the system.
  \item For complete spheres, $r \ge 0$, $\rho(r)$ and $m(r)$ are in 1-1 correspondence thanks to (\ref{density}). Either is equivalent to specifying $e^{\pm 2\Psi(r)}$. In the presence of a core, $\rho(r)$ determines $m(r)$ up to an arbitrary constant, the unspecified mass of the core. This point will be elaborated in Sections \ref{sec:42} as we will revisit the existing algorithms.
  \item In both (\ref{momentum_conservation}) and (\ref{EFE}), only $\Phi'(r)$ appears. The additive constant resulting from the integration of $\Phi(r)$ becomes a multiplicative constant in the line element and, as such, can be absorbed into the time coordinate of (\ref{line_element}). Thus, we can treat $\Phi'(r)$  as the basic function instead of $\Phi(r)$.
\end{enumerate}

\subsection{Algorithm for Generating All Isotropic Solutions}\label{sec:41}
The isotropic system needs one input to generate a solution. This could be an equation of state, one of $\Phi(r)$, $\rho(r)$, and $p(r)$, or any suitable combination of these. Exact solutions of this system have been studied extensively and were mostly obtained using \emph{ad hoc} assumptions on $\rho(r)$, $e^{2\Phi(r)}$ (equivalently, $\Phi'(r)$), and $e^{-2\Psi(r)}$ (equivalently, $m(r)$). Out of the three basic functions, only one can serve as a generating function, as we will see below:
\begin{itemize}
  \item $\rho(r)$: If one uses $\rho(r)$ as the input, the TOV equation (\ref{generalized_TOV1}) is a (first-order) Riccati equation in $p(r)$ of the form $p'+f(r)p+g(r)p^2=h(r)$. By replacing $p(r)$ in (\ref{generalized_TOV1}) and using (\ref{momentum_conservation}), on the other hand one gets a Riccati equation in $\Phi'(r)$. In either case, the basic problem is that the general Riccati is not solvable by quadratures. Thus, one cannot create an algorithm to generate all solutions using $\rho(r)$ as the input.
  \item $p(r)$: With $p(r)$ as the input, the TOV equation is an Abel equation of the second kind in $m(r)$ (see, for example, \cite{zwillinger1998handbook}). This is a more difficult equation to solve than the Riccati equation, even with special choices of coefficients, which perhaps explains why there is not a single example of $p(r)$ being used as the input to obtain an exact solution for the isotropic system \cite{kramer1980exact}.
  \item $\Phi'(r)$: With $\Phi'(r)$ as input, the TOV equation turns into a first-order linear differential equation in $m(r)$, which, as Weynman noticed in 1949, ``can always be solved by quadratures" \cite{Wyman:1949zz} and thus, one can have an algorithm that generates all solutions \cite{PhysRevD.67.104015}. The details of this algorithm are contained in the corresponding anisotropic algorithm discussed below. This has, thus, been a popular route for obtaining an exact solution, too \cite{kramer1980exact}. However, it is not easy to know beforehand what choice of $\Phi(r)$ would produce the desired properties of $\rho(r)$ and $p(r)$ or a particular equation of state. To generate solutions that are regular at the center, $\Phi(r)$ must have the property that $\Phi'(0)=0$ and $\Phi''(0)$ must be finite, as a necessary condition, as we discussed above in Section \ref{sec:iii} and in \cite{PhysRevD.67.104015}. We will address this below and see how Einstein's equations will fill the gap for regularity.
\end{itemize}

\subsection{Generating Regular Anisotropic Solutions from the Existing Algorithm}\label{sec:42}

For the two simultaneous inputs needed for the anisotropic system, if one combines anisotropy $\chi(r)$ (or, equivalently, $P(r)$) with $\rho(r)$, $p(r)$, or $\Phi(r)$, the qualitative behaviors remain the same as above. For $(\rho(r), P(r))$ and $(\Phi(r), P(r))$, the generalized TOV is still a Riccati equation and an Abel equation of the second kind in $m(r)$, respectively, with no gain in their general solvability. The only generating pair is $(\Phi(r), P(r))$, which a natural extension of the isotropic case, and has been worked out using $\chi(r)$ \cite{herrera2008all}. Below we will rederive this and look for specialized algorithms that generate all regular solutions.

\subsubsection*{Existing Algorithm for (An)isotropic Solutions: A Second Look}
The generalized TOV can be turned into a first-order differential equation in $m(r)$ for either $(\Phi(r), P(r))$ or $(\Phi(r), \chi(r))$ as input. One can solve (\ref{EFE}) algebraically for $p(r)$ and substitute it in (\ref{generalized_TOV1}) or (\ref{generalized_TOV11}) with $m'(r)=4\pi\rho r^2$ to get\footnote{One can also use $G^\theta_\theta$ and $G^r_r-G^\theta_\theta$ equations for this.}:
\begin{equation}
\label{firstorderode}
\frac{dm(r)}{dr}+ a(r) m(r)=b(r).
\end{equation}
Unlike the Riccati or the Abel equation, this can be solved exactly for arbitrary forms of its variable coefficients:
\begin{equation}
\label{masssolutionsherraraincano}
m(r)=\frac{\int b(r) e^{\int a(r) dr} dr + C}{e^{\int a(r) dr}}.
\end{equation}
The coefficients for $(\Phi(r), P(r))$ are
\begin{align}
\label{differentalgebraicexpressions}
a(r) &= \left(\frac{2r^2(\Phi'^2+\Phi'')+r\Phi'-1}{r(r\Phi'+1)}\right),\\
\label{differentalgebraicexpressions2}
b(r) &= \frac{r(\Phi'+r\Phi''+r\Phi'^2-8\pi rP)}{(r\Phi'+1)}.
\end{align}
Thus, there is a one-parameter family of solutions $m(r)$ for any input $(\Phi(r), P(r)$), which will give a parameter-dependent $p(r)$ algebraically via (\ref{EFE}). The resulting one-parameter family of geometries found by substituting (\ref{masssolutionsherraraincano}) and $\Phi(r)$ in the line element (\ref{line_element})  will differ nontrivially in its $g_{rr}$ component for different values of $C$. The isotropic case is recovered by setting  $P(r)=p(r)$; in this case, one would have eliminated at the previous step the $P(r)$ appearing in $b(r)$. If one uses $\chi(r)$ instead of $P(r)$, the coefficients of (\ref{firstorderode}) are then as follows:
\begin{align}
\label{differentalgebraicexpressions1}
a(r) &= \left(\frac{2r(\Phi''+\Phi'^2)}{(r\Phi'+1)}-\frac{3}{r}\right),\\
\label{differentalgebraicexpressions11}
b(r) &= \frac{r(r\Phi''+r\Phi'^2-\Phi'-24\pi r\chi)}{(r\Phi'+1)}.
\end{align}
The algorithm presented in \cite{herrera2008all} uses the inputs $(z(r), \Pi(r))$:
\begin{align}
\label{Herraravariables}
z(r) &= \Phi'+1/r,\\
\Pi  &= 8\pi(p-P),
\end{align}
which, from the discussion at the beginning of this section, are equivalent to the $(\Phi'(r), P(r))$ or $(\Phi'(r), \chi(r))$ that we used above. It also used $y(r)=1-2m(r)/r$ instead of $m(r)$. This is $e^{-2\Psi(r)}$. With these, it is easy to check that (\ref{firstorderode}) is precisely equation (8) of \cite{herrera2008all}:
\begin{equation}
\label{Herrer}
y'+y\left(\frac{2z'}{z}+2z-\frac{6}{r}+\frac{4}{zr^2} \right)=-\frac{2}{z} \left(\Pi+\frac{1}{r^2}\right).
\end{equation}
We do not reproduce the final one-parameter $y(r)$ or the metric expressed in terms of $z$ and $\Pi$ \cite{herrera2008all}. Although completely equivalent, using $m(r)$ instead of $y(r)$ makes it slightly easier to interpret the free parameter in the general solution.

\subsubsection*{Regular Solutions using $\Phi(r)$ and $P(r)$}
There is no discussion in the literature on the role of the parameter that appears in the solution of $y(r)$ in \cite{herrera2008all} or in $m(r)$ in \cite{PhysRevD.67.104015} for the isotropic case. We will now identify the parameter by looking at (\ref{firstorderode}) from the point of view of an initial-value problem first and discuss its subsequent role as $r\to 0$. Assuming, there is a core, i.e., $r=r_c> 0$, one needs to solve the following initial-value problem for $r\ge r_c$:
\begin{equation}
\label{firstorderodeivp}
\frac{dm(r)}{dr}+ a(r) m(r)=b(r), \quad  m(r_c)=m_c,
\end{equation}
where $m_c$ is the mass of the core. This has the unique solution:
\begin{equation}
\label{generalsolmass}
m(r)= \frac{\int_{r{_c}}^r b(x) e^{\int_{r{_c}}^x a(s) ds} dx+ m_c  }{e^{\int_{r{_c}}^r a(s) ds}},
\end{equation}
which, on comparison with (\ref{masssolutionsherraraincano}), shows that $C= m_c$
precisely. Thus, the parameter $C$ in our formulation above, and in that of \cite{PhysRevD.67.104015}, represents the mass of the (nonzero) core.\footnote{Note that this does not follow from the look of  (\ref{firstorderode}) since $C$ is in its numerator.} (We will soon see that for complete spheres it will assume a different meaning.) One can further qualify $C$ on physical grounds for nonzero cores. For example, for a Schwarzschild core, one must have:
\begin{equation}
m_c\le \frac{r_c}{2}.
\end{equation}
Thus, contrary to the impression given by the general solution (\ref{masssolutionsherraraincano}) of $m(r)$, not all possible values of the parameter $C$ are admissible.

For complete fluid spheres with $r\ge 0$, unfortunately, one cannot set $m(0)=0$ to get $C=0$ as $r=0$ is a singular point of the differential equation ($\lim_{r\to 0^{+}} a(r)$ diverges). However, $r=0$ is a regular singular point since $\lim_{r\to 0^{+}} ra(r)$ is finite (see, for example, \cite{CoddingtonLevinson}). To analyze the solution near it, we consider the dominant terms in $a(r)$ and $b(r)$ as $r\to 0$:
\begin{align}
\label{differentalgebraicexpressionsleading}
a(r) &= -\frac{1}{r},\\
b(r) &= r^2\left[2\Phi''(0)-8\pi P(0)\right],
\end{align}
where we have used $\lim_{r\to 0^+}\Phi(r)/r=\Phi''(0)$. This yields the following general solution:
\begin{equation}
\label{leadingordermasswithP}
m(r)=\left[\Phi''(0)-4\pi P(0)\right] r^3+ C_1 r.
\end{equation}
Imposing the initial value $m(0)=0$ will not allow us to determine the arbitrary constant $C_1$, illustrating the breakdown of the uniqueness theorem at the singular point $r=0$. This linear term will prevent $\lim_{r\to 0^+} m(r)/r^3$ from being finite and will give a non-regular solution for any $C_1\ne 0$.\footnote{The condition $m(0)=0$ is a necessary but not sufficient condition of regularity; it avoids a conical singularity \cite{Wald:1984rg} and one needs $m(r)/r^3$ to be finite as we discussed earlier.} This is expected from our discussions on regularity in Section \ref{sec:iii} since $\Phi'(0)=0$, $\Phi''(0)$ finite, and $P(0)$ finite are not sufficient for regularity.

The cubic term in (\ref{leadingordermasswithP}) comes from the first term of (\ref{masssolutionsherraraincano}). The linear term, on the other hand, comes from the second term of (\ref{masssolutionsherraraincano}), $Ce^{-\int a(r) dr}$, which gives the difference between any two solutions of (\ref{firstorderode}). It is, thus, easy to see that $C_1=C$ in~(\ref{leadingordermasswithP}). So, if we can find a condition that requires $C_1=0$ in (\ref{leadingordermasswithP}), we can set $C=0$ in general.

Fortunately, this can be argued from the general solution (\ref{masssolutionsherraraincano}) as a limiting process as follows. After integrating the diverging term $1/r$, (\ref{masssolutionsherraraincano}) can be rewritten as
\begin{equation}
	\label{general_solution_density}
	\frac{m}{r^3}=\frac{C}{r^2}\,e^{-A(r)}+\frac{e^{-A(r)}}{r^2}\left(\int \frac{b(r)}{r}e^{A(r)}dr \right)
\end{equation}
where
\begin{equation}
	A(r)=\int \alpha(r)dr= 2 \int \frac{\Phi'+r(\Phi''+\Phi'^2)}{(r\Phi'+1)} dr.
\end{equation}
It is easy to see as $r\to 0$, both $\alpha(r)$ and $b(r)/r$ go to zero. As $r\to 0$ the first term of the right hand side diverges since the integrand $\alpha(r)\to 0$.  However, the other part is finite:
\begin{equation}
	\lim_{r\to 0}\left(\int \frac{b(r)}{r}e^{A(r)}dr \right)/r^2=\Phi''(0)-4\pi P(0).
\end{equation}
Since $\lim_{r\to 0^+} m/r^3= 4\pi\rho(0)/3$, one get from (\ref{general_solution_density})
\begin{equation}
	\frac{4\pi}{3}\rho(0)=\Phi''(0)-4\pi P(0).
\end{equation}
This is the regularity condition (\ref{algebraicrelationP}). We, thus, have the following theorem.
\\
\\
\textbf{Theorem 4.1:} For a differentiable function $\Phi'(r)$ and continuous function $P(r)$ defined on the common interval $[0, r_b)$, $0 <r_b <\infty$, such that $\Phi'(0)=0$ and $\Phi''(0)$ is finite, and $P(0)$ is finite, the input $(\Phi'(r), P(r)$)
generates a unique anisotropic fluid sphere that is regular at the center with mass given by
\begin{equation}
\label{massHerraraagain}
m(r)=re^{-\int_{0}^{r}\alpha(s)ds}\int_{0}^{r}\frac{b(x)}{x}e^{\int_{0}^{x}\alpha(s)ds} dx,
\end{equation}
and central density given by
\[
\rho(0)=\frac{3}{4\pi}\Phi''(0)-3P(0).
\]
\subsubsection*{Regular Solutions using $\Phi(r)$ and $\chi(r)$}
The above (i.e., the vanishing of $C_1$ and, hence, $C$) can also be argued by invoking the vanishing of anisotropy at the center in regular solutions, i.e., $\chi(0)=0$. Using anisotropy, i.e., (\ref{differentalgebraicexpressions1}) and (\ref{differentalgebraicexpressions11}) for $a(r)$ and $b(r)$, and keeping the dominant terms as before one gets:
\begin{align}
\label{differentalgebraicexpressionsleading}
a(r) &= -\frac{3}{r},\\
b(r) &= r^4\Phi''(0),
\end{align}
which gives
\begin{equation}
\label{leadingordermasswithchi}
m(r)= C r^3 + \frac{1}{2}\Phi''(0)r^5.
\end{equation}
Here, opposite to (\ref{leadingordermasswithP}), the first and second terms come, respectively, from the second and first terms of the corresponding general solution (\ref{masssolutionsherraraincano}).\footnote{One would get an identical result using (\ref{Herrer}).} Other terms in the expansions of $\Phi'(r)$ and $\chi(r)$ would produce terms with powers higher than three, as one can easily check.

Comparing (\ref{leadingordermasswithP}) with (\ref{leadingordermasswithchi}) clearly shows that we have to set $C_1=0$ in (\ref{leadingordermasswithP}), and hence, $C=0$ in (\ref{masssolutionsherraraincano}), to achieve isotropy and regularity simultaneously at the center. This also shows that in (\ref{leadingordermasswithchi}) $C=\Phi''(0)-4\pi P(0)$, which could not be achieved using only anisotropy. The resulting $m(r)$ and $p(r)$ will be parameter-free and so the metric (\ref{line_element}) will be unique for every appropriate choice of $(\Phi'(r), P(r)$).

The general solution (\ref{masssolutionsherraraincano}) with $\chi(r)$ can also be used for $r=0$ provided one imposes the conditions of regularity as $r\to 0^+$. This can be seen by rewriting the general solution as
\begin{equation}
\label{herreraalgoreg}
\frac{m}{r^3}=e^{-A(r)}C+e^{-A(r)}\left(\int \frac{b(r)}{r^3} e^{A(r)} dr\right)
\end{equation}
where
\begin{equation}
	A(r)=\int \alpha(r)dr=\int \frac{2r(\Phi''+\Phi'^2)}{(r\Phi'+1)} dr.
\end{equation}
It is easy to check that $\lim_{r\to 0^+}\alpha(r)=0$ and $\lim_{r\to 0^+}b(r)/r^3$ is finite if, for $\chi(0)=0$, $\chi'(0)$ and $\Phi'''(0)$ are finite. However, these extra conditions do not appear in the Einstein equations for the system. So, the right way to recover the solution at $r=0$ from the general form (\ref{herreraalgoreg}) above (which was obtained for $r \ne 0$) is, first, to impose the conditions of regularity: $\chi(0)=0$ and $\lim_{r\to 0^+}\Phi(r)/r=\Phi''(0)$. After integration, the second term in (\ref{herreraalgoreg}) will then precisely reproduce the $\Phi''(0)r^5/2$ term of (\ref{leadingordermasswithchi}). With this, (\ref{herreraalgoreg}) is a regular one-parameter family of solutions.\footnote{For the isotropic system, this freedom with the parameter is what was used in \textbf{Theorem (P2)} in \cite{boonserm2007solution} to generate new solutions (with $C=4\pi(\delta\rho)/3$ where $\delta\rho$ is shift in central density).}
\\
\\
\textbf{Theorem 4.2:} For two differentiable functions $\Phi'(r)$ and $\chi(r)$ defined on the common interval $[0, r_b)$, $0 <r_b <\infty$, such that $\Phi'(0)=\chi(0)=0$ and $\Phi''(0)$ is finite, each input $(\Phi'(r), \chi(r)$) generates a one-parameter family of anisotropic fluid spheres that are regular at the center.
\\
\\
Thus, the conditions $\Phi'(0)=\chi(0)=0$ and finite $\Phi''(0)$ are sufficient conditions for regularity for anisotropic fluid spheres. Note that these conditions alone do not make $\kappa$ finite. However, they prove sufficient for the system because the Einstein equations lead to $m(r)$ having the right behavior under these conditions, making $\kappa$ finite. We, thus, have the following for the isotropic system:
\\
\\
\textbf{Corollary 4.3:} For a differentiable function $\Phi'(r)$ defined on the interval $[0, r_b)$, $0 <r_b <\infty$, such that $\Phi'(0)=0$ and $\Phi''(0)$ is finite, the input function $\Phi'(r)$ generates a one-parameter family of isotropic fluid spheres that are regular at the center.
\subsection{Three New Choices and an Algorithm}\label{sec:43}
We now consider the remaining three combinations of the four variables of the anisotropic system: $(\rho(r)$, $p(r)$), $(\rho(r),\Phi(r))$, and $(p(r), \Phi(r))$. The line element (\ref{line_element}) can be formally integrated and be expressed it in terms of $\rho(r)$ and $p(r)$\footnote{This is obtained by rewriting $\Phi(r)$:
\begin{align}
\label{EFE1}
\frac{d\Phi}{dr} &= \frac{m(r)+4\pi r^3 p}{r(r-2m(r))} \\
 &= \frac{m(r)}{r(r-2m(r))}+ \frac{4\pi r^3 p(r)}{r(r-2m(r))}\\
 &= \frac{1}{2} \left[\frac{1}{r-2m(r)}-\frac{1}{r}\right] +\frac{4\pi r^2 p(r)}{r-2m(r)}\\
 &= \frac{1}{2} \left[\frac{1-2m'(r)}{r-2m(r)}-\frac{1}{r}\right] +\frac{4\pi r^2 p(r)}{r-2m(r)}+\frac{m'(r)}{r-2m(r)}.
\end{align}}:
\begin{equation}
\label{line_elementsupersimplified}
ds^2= -\left(1-\frac{2m(r)}{r}\right){e^{8\pi\int \frac{\left(\rho(r)+p(r)\right) r^2}{r-2m(r)}\, dr}} dt^2+ \frac{dr^2}{1-\frac{2m(r)}{r}} +r^2 (d\theta^2+\sin^2{\theta} d\phi^2).
\end{equation}
This metric requires two integrations to generate any particular solution, but it is now abundantly clear that one can generate \emph{all} solutions by specifying $(\rho(r), p(r))$. However, what about the other two pairs and regular solutions? The answer can be summarized as follows.
\\
\\
\textbf{Lemma 4.4:} For the anisotropic system of fluids in canonical coordinates, any pair of the input functions  $(\rho(r)$, $p(r)$), $(\rho(r),\Phi(r))$, or $(p(r), \Phi(r))$:
\begin{enumerate}
	\item[(a)] are equivalent,
  \item[(b)] generate all solutions,
  \item[(c)] determine the geometry (\ref{line_element}) uniquely upon two integrations,
  \item[(d)] determine $P(r)$ uniquely.
\end{enumerate}
\textbf{Proof:} We first note that the algebraic relation (\ref{momentum_conservation}) is linear in $\Phi'(r)$, $m(r)$, and $p(r)$. Thus, any two determine the third one uniquely.\footnote{Explicitly:
\begin{align}
\label{differentalgebraicexpressions}
\Phi'(r) &= \frac{m(r)+4\pi r^3 p}{r(r-2m(r))},\\
p(r)     &= \frac{r(r-2m(r))\Phi'(r)-m(r)}{4\pi r^3},\\
m(r)     &= \frac{r^2\Phi'(r)-4\pi r^3 p(r)}{1+2r\Phi'(r)}.
\end{align}}
We have already noted that $\Phi'(r)$ and $\Phi(r)$ are equivalent and that $\rho(r)$ and $m(r)$ are equivalent if we have the origin at $r=0$ (with zero cental mass) or a core $r\le r_c$ with a known mass. Thus, $(m(r)$, $p(r)$), $(\rho(r),\Phi(r))$, $(p(r), \Phi(r))$, and $(\rho(r)$, $p(r)$) are equivalent inputs. In the absence of any such boundary condition, purely at the differential equation level, specifying $\rho(r)$ will return a family of functions $m(r)$ with an arbitrary additive constant (parameter) $C$. The equivalence will continue to hold, since ($\rho(r), p(r))$ and ($\rho(r), \Phi(r))$ will then, respectively, determine $\Phi(r)$ and $p(r)$ via the same algebraic relation above, up to the same parameter. On the other hand, using $(p(r), \Phi(r))$ as input will determine $m(r)$ and hence, $\rho(r)$ uniquely.

Given that $(\rho(r)$, $p(r)$), $(\rho(r),\Phi(r))$, and $(p(r), \Phi(r))$ are equivalent inputs, it is sufficient to work with the pair $(\rho(r)$, $p(r)$) to see that it generates all solutions. $\rho(r)$ determines $m(r)$ and hence, $g_{rr}$ and the pair $(\rho(r)$, $p(r)$) fixes $\Phi(r)$ and hence, $g_{tt}$ uniquely and determines $P(r)$ uniquely algebraically via (\ref{generalized_TOV1}). Thus, for any of $(\rho(r)$, $p(r)$), $(\rho(r),\Phi(r))$, or $(p(r), \Phi(r))$, one requires two integrations to find $m(r)$ and $\Phi(r)$.
For fluid solutions starting from $r=0$ or surrounding a central core of known mass, every choice of input pair will generate a unique sphere. On the other hand, if one does not consider a boundary value imposed by the center, as we discussed above, one gets a one-parameter family.

It is important to appreciate that nothing trivial is happening here. The line element (\ref{line_elementsupersimplified}) is equally valid for the isotropic case, but there $\rho(r)$ and $p(r)$ cannot be specified simultaneously. For the same reason, the algebraic equivalence of the three pairs loses its significance (except at the level of regularity) in the isotropic case, since one can specify only one input. On the other hand, specifying $\rho(r)$ or $p(r)$ leads to Abel or Riccati equations, which are very difficult to solve in general. Here, we can choose $\rho(r)$ and $p(r)$ simultaneously and independently, which makes it immensely simpler to satisfy the regularity conditions. This is a clear advantage of the anisotropic system over the isotropic system.

The isotropic limit of any anisotropic solution with the same potential function can be found by applying the algorithm of \cite{PhysRevD.67.104015}. For a given potential function $\Phi(r)$, if $m_A(r)$ and $m_I(r)$ are the mass functions for the anisotropic and isotropic solutions, respectively, then by introducing a parameter $\alpha$, one can write:

\begin{equation}
\label{isotropic_limit}
m_A(r)=m_I(r)+\alpha F(r).
\end{equation}

Since the field equations (\ref{GrrEinstein}) and (\ref{EEthetatheta}) are linear (as an algebraic equation and as a differential equation) in the mass function, the radial and tangential pressures have linear relations with the isotropic pressure. (See \cite{Casadio:2019usg} or substitute (\ref{isotropic_limit}) into (\ref{GrrEinstein}) and (\ref{EEthetatheta}) to obtain $p(r)$ and $P(r)$ in terms of the isotropic pressure and $F(r)$.)

\subsubsection*{Regular Solutions via the New Algorithm}
Combining with the discussions on regularity in Section  \ref{sec:iii}, it is very clear how one can generate all regular anisotropic solutions using the above algorithm in addition to Theorem 4.1
\\
\\
\textbf{Theorem 4.5:}
All anisotropic fluid solutions of Einstein equations in canonical coordinates that are regular at the center can be generated using any of the following:
\begin{enumerate}
	\item[(a)] $(\rho(r)$, $p(r)$) with $\rho(0)$ finite, $p(0)$ finite, and $\lim_{r\to 0} r p'(r)=0$,
  \item[(b)] $(\rho(r), \Phi(r))$ with $\rho(0)$ finite, $\Phi'(0)=0$, and $\Phi''(0)$ finite,
  \item[(c)] $(p(r), \Phi(r))$ with $p(0)$ finite, $\lim_{r\to 0} r p'(r)=0$, $\Phi'(0)=0$, and $\Phi''(0)$ finite.
\end{enumerate}
This algorithm cannot be specialized to produce only isotropic solutions. However, it does generate all isotropic solutions as the special case when the pair $(\rho(r), p(r))$ satisfies the TOV equation of the isotropic system.

\section{Harvesting the Riccati}\label{sec:v}
An interesting feature of a Riccati equation is that if a particular solution is known, one can find the general solution by adding to the particular solution the general solution of an associated Bernoulli equation.\footnote{There is a nice discussion in the appendix of \cite{boonserm2007solution} on the possible forms of general solutions of the Riccati equation when one, two, or three particular solutions are known; in all cases there is one single constant of integration, as expected. There are a number of transformations that convert a general Riccati into a homogeneous linear second-order ordinary differential equations whose general solutions have two arbitrary constants; however, all of these map back to a one-parameter solution of the Riccati equation.} Finding a particular solution, thus, is as difficult as finding the general solution, unless the particular solution is exported from elsewhere. The two algorithms can do just that.
\subsection*{Riccati in $p(r)$}
Any ``seed" solution $(\rho_i(r)$, $p_i(r), \Phi'_i(r), \chi_i(r))$ or, equivalently, $(\rho_i(r)$, $p_i(r), \Phi'_i(r), P_i(r))$ generated by either algorithm can be seen as a particular solution of equation (\ref{generalized_TOV1}) and (\ref{generalized_TOV1}). They both are Riccati equations in $p(r)$:
\begin{equation}
\label{generalized_TOV}
\frac{dp}{dr}+f(r)p+g(r)p^2=h(r),
\end{equation}
where the coefficients $f(r)$, $g(r)$, and $h(r)$ are determined by $\rho_i(r)$ and $\chi_i(r)$ as in (\ref{generalized_TOV11}) or by $\rho_i(r)$ and $P_i(r)$ as in (\ref{generalized_TOV1}). The general solution of (\ref{generalized_TOV11}) is
\begin{equation}
\label{RiccatiGeneralSol}
p_N(r)=p_{i}(r)+
\frac{C e^{-\int \left( 2g(r) p_{i}(r)+f(r)\right)dr}}{1+C \int g(r) e^{-\int \left(2 g(r) p_{i}(r)+f(r)\right)dr}dr}.
\end{equation}
It is easy to check that as $r\rightarrow 0$, both $f(r)\rightarrow 0$ and $g(r)\rightarrow 0$ under the same conditions of regularity discussed in Section \ref{sec:iii}. Thus,
\begin{equation}
\label{RiccatiGeneralSolcenter}
p_N (0)=p_{i} (0)+C.
\end{equation}
However, since anisotropy has been kept the same, and thus still vanishes at the center, in the new solution the two pressures increase or decrease by the same amount $C$ and that the new solution is regular for any $C$. Thus one gets a one-parameter family of regular solutions for any regular seed solution.

On the other hand, the general solution of (\ref{generalized_TOV1}), where $f(r)$, $g(r)$, and $h(r)$ are determined by $\rho_i(r)$ and $P_i(r)$, the general solution is
\begin{equation}
	p(r)=p_{i}(r)+\frac{C e^{-\int \left( 2g(r) p_{i}(r)+f_0(r)\right)dr}}{r^2\left(1+C \int {g(r) \over r^2} e^{-\int \left(2g(r) p_{i}(r)+f_0(r)\right)dr}dr\right)},
\end{equation}
where $f_0(r)=f(r)-2/r$. It is easy to see that unless the second term vanishes, there is no way one can get $p(r)=P_i(r)$. So there will be nonzero anisotropy at the center and the one-parameter solution is not regular.
\subsection*{Riccati in $\Phi'(r)$}
As we mentioned earlier, the generalized TOV can also be seen as a Riccati equation in $\Phi'(r)$,
\begin{equation}
\Phi''(r)+\left[\frac{r-m_i-4\pi r^3\rho_i(r)}{r(r-2m_i)}\right] \Phi'(r)+ \Phi'^2(r)=\frac{8\pi r^3 P_i(r)+4\pi r^3\rho_i(r) -m_i}{r^2(r-2m_i)}.
\end{equation}
One can thus generate another one-parameter solution using a regular seed solution and it is not difficult to see that
\begin{equation}
\label{RiccatiGeneralSolcenterphi}
\Phi'_N (0)=\Phi_i' (0)
\end{equation}
and thus the resulting one-parameter solutions can be regular. However, since this uses the same input $(\rho(r), P(r))$ as the Riccati equation in $p(r)$, all solution of the latter can be combined with $\rho(r)$ algebraically via (\ref{EFE}) to give the same $\Phi'(r)$ and hence one will get the same one-parameter family of solutions. We, thus, have the following theorem:
\\
\\
\textbf{Theorem 5.1:}
For any (an)isotropic fluid solution of the Einstein equations in canonical coordinates that is regular at the center, one can generate a one-parameter family of regular (an)isotropic solutions by solving the (generalized) TOV equation as Riccati equations in $p(r)$ or $\Phi' (r)$.
\\\\
This would apply to all solutions including those generated using equations of state. The resulting one-parameter solutions, however, will not in general obey the equation of state of the seed.

\section{Physical Solutions: Two New Classes of Anisotropic Solutions} \label{sec:vi}
In addition to central regularity and matching with the Schwarzschild metric at the boundary, the following properties are often sought in a physical fluid sphere (see, for example, \cite{delgaty1998physical,Harko:2002db}):
\begin{enumerate}
	\item $\rho(r)$, $p(r)$ and $P(r)$ are positive and monotonically decreasing. \label{physical_condition_1}
	\item The solution is regular at the center. \label{physical_condition_2}
	\item $p(r)$ vanishes at the boundary.\footnote{$P(r_b)$ is not necessarily required to vanish.}
	\item Fluid variables satisfy the energy conditions: $\rho+p+2P\geq 0$ and $\rho\geq p+2P$.
	\item Speed of sound is less than speed of light, i.e., $0\leq dp/d\rho \leq 1$ and $0\leq dP/d\rho \leq 1$.\footnote{Actually, the quantities, $dp/d\rho$ and $dP/d\rho$, don't necessarily represent sound speed; additional significant assumptions (e.g., barotropic equation of state or adiabatic star) are required to enforce the relationship, $0\leq dp/d\rho \leq 1$ and $0\leq dP/d\rho \leq 1$ (see, for example, section $3.8$ of \cite{Rahman:2001hp}).} \label{sound_speed}
\end{enumerate}
Apart from these conditions, it is required that the metric component $g_{rr}$ is positive (i.e., $r>2m$).
However, this is a consequence of conditions (\ref{physical_condition_1}) and (\ref{physical_condition_2}) (see, for example, \cite{Rendall_1993}, where it was shown, through regularity analysis of the Riccati equation (\ref{generalized_TOV1}), that for positive density, differentiable tangential pressure and finite isotropic central pressure, $r>2m$).

\subsection*{Example I}
Consider the following density and radial pressure as two input functions:
\begin{equation}
  \rho(r)=\alpha\left(r_b^2-r^2\right),\quad p(r)=\beta\left(r_b^2-r^2\right)^2,\quad \alpha,\beta>0,\quad r\in [0,r_b],
\end{equation}
where $r_b>0$ is a boundary.
Thus, the radial pressure and density follow the relationship, $p\propto\rho^2$, and vanish simultaneously at the boundary.
Furthermore,
\begin{equation}
    \rho'(r)=-2\alpha\,r<0,\quad p'(r)=-4\beta\,r\left(r_b^2-r^2\right)<0.
\end{equation}
From condition (\ref{correctinpus2}), the solution is regular the center.
The mass function is given by
\begin{equation}
  \label{Example_1_mass}
  m(r)=\left(\frac{4\pi\alpha\, r_b^2}{3}\right)r^3-\left(\frac{4\pi \alpha}{5}\right)r^5.
\end{equation}
Thus, the mass function is same as in the Tolman-VII solution \footnote{This solution is one of $16$ isotropic solutions (out of $127$) which satisfy all the physical conditions listed above \cite{delgaty1998physical}.} \cite{Tolman:1939jz}.
It turns out that the tangential pressure vanishes at the boundary as well and is given by
\begin{equation}
  P(r)=\frac{\left(r^2-r_b^2\right)\left[9\beta\,r^2\left(5+4\pi \alpha r^4\right)-\beta\,r_b^2\left(15+68\pi \alpha r^4\right)+2\pi \alpha^2 r^2\left(3r^2-5r_b^2\right)+30\pi \beta^2\,r^2 \left(r^2-r_b^2\right)^3 \right]}{15+8\pi \alpha\,r^2\left(3r^2-5r_b^2\right)}.
\end{equation}
Furthermore, the $g_{tt}$ component of metric, from (\ref{line_elementsupersimplified}) reduces to
\begin{equation}
  g_{tt}=-\left(1-\frac{2m}{r}\right)^{-\xi} \exp\left[\frac{5\beta}{2\alpha}\left(r^2-\frac{5}{6}r_b^2\right)+\frac{5}{12\alpha\delta}\left(\frac{\beta r_b^4}{6}+\alpha r_b^2-6\beta\delta^2\right)\tan^{-1}\left(\frac{6r^2-5r_b^2}{6\delta}\right)\right]
\end{equation}
where
\begin{equation}
	\xi=\frac{1}{4}+\frac{5\beta r_b^2}{12\alpha}\quad \text{and}\quad \delta^2=\frac{5}{8\pi\alpha}-\frac{25}{36}\;r_b^4\,>0\quad\text{i.e.,}\quad \alpha<\frac{9}{10\pi r_b^4}.
\end{equation}

\begin{figure}[h!]
  \makebox[\textwidth][c]{\includegraphics[width=1.15\textwidth]{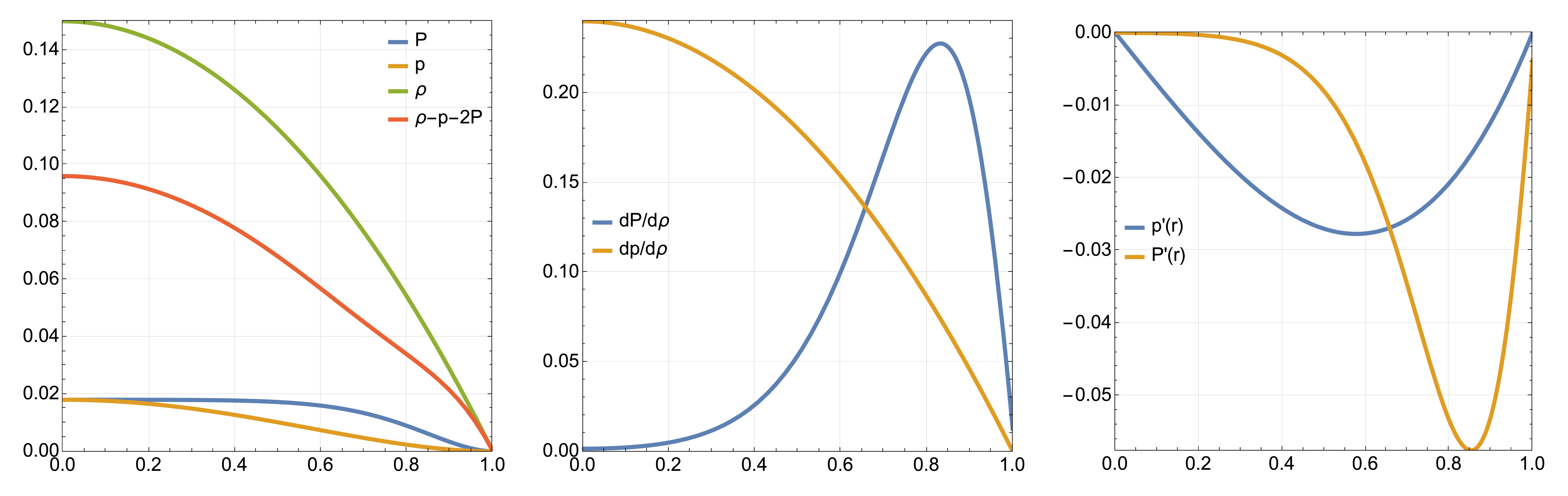}}
  \caption{
  Here, $r_b=1$, $\alpha=0.15$, and $\beta=0.018$. Thus, the Schwarzschild mass is $M=0.25$ and radius is $R=3.9\,M$. As can be seen, this solution satisfy all the physical conditions.}
\end{figure}

\begin{figure}[h!]
  \makebox[\textwidth][c]{\includegraphics[width=1.15\textwidth]{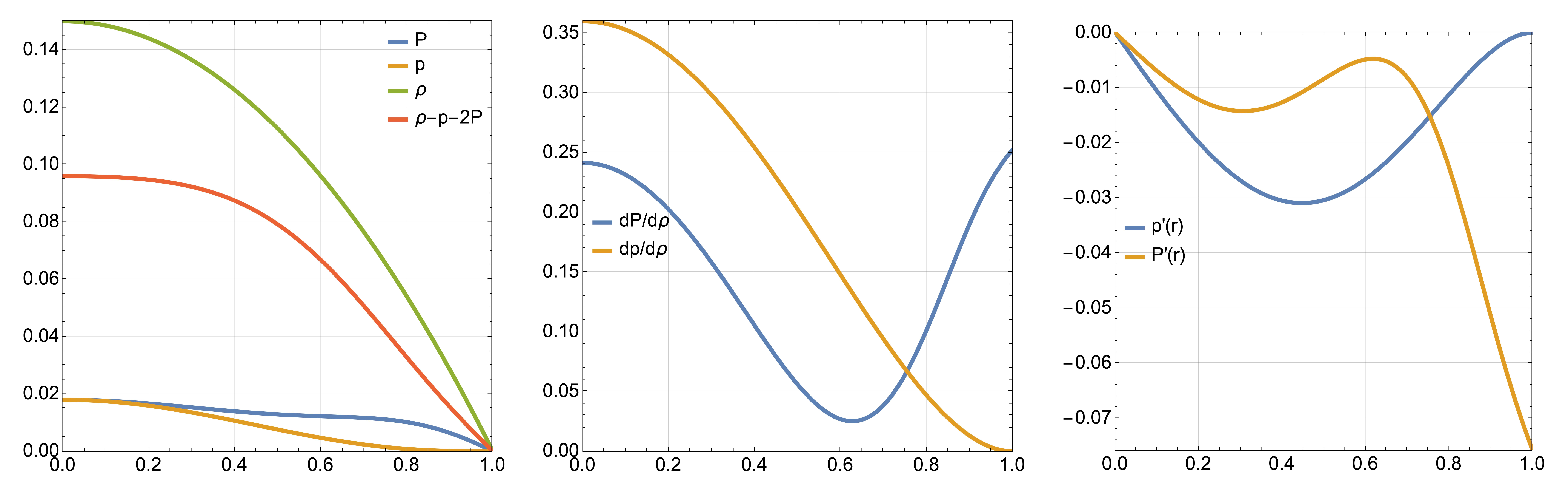}}
  \caption{If the relationship between the radial pressure and density is changed to $p\propto \rho^3$, while $\rho(r)$ is still the same, then the new solution still satisfy all the physical conditions. Here, $r_b=1$, $\alpha=0.15$ and $\beta=0.018$.}
\end{figure}

\subsection*{Example II}
Consider the following potential function and density profile as two input functions:\footnote{
Parametrizing this way yields $\Phi'(r)>0$, recall equation (\ref{EFE}).}
\begin{equation}
	\label{two_inputs}
	\Phi(r)=\gamma^2 r^2,\quad \rho(r)=\rho_0-\left(\rho_0-\rho_b\right)\left(\frac{r}{r_b}\right)^2, \quad 0\leq \rho_b <\rho_0, \quad r\in [0,r_b].
\end{equation}
In other words, the $g_{tt}$ component of the metric is Gaussian.
Moreover,
\begin{equation}
	\Phi'(0)=0,\quad \Phi''(r)=2\gamma^2,\quad \rho(0)=\rho_0,\quad \rho(r_b)=\rho_b.
\end{equation}
Therefore, from (\ref{correctinpus0}), the solution is regular at the center. Since
\begin{equation}
	\rho'(r)=-\frac{2r}{r_b}(\rho_0-\rho_b) \leq 0,
\end{equation}
density is monotonically decreasing and $\rho(r)\geq \rho(r_b)\geq 0$.
We will apply the boundary condition $p(r_b)=0$ to determine $\rho_b$ in terms of $\rho_0$ and $r_b$.
The second boundary condition $\exp[{2\Phi(r_b)}]=1-(2m(r_b)/r_b)$ can be satisfied by rescaling the time coordinate with appropriate constant.
In other words, the two parameters one can choose in this approach are $\rho_0$ and $r_b$.
The mass function reduces to
\begin{equation}
  m(r)=\frac{4\pi\rho_0r^3}{3}-\frac{4\pi r^5}{5} \left(\frac{\rho_0-\rho_b}{r_b^2}\right)=\frac{r^3}{a^2}-\frac{r^5}{b^2}.
\end{equation}
Once again, the mass function is the same as that in the Tolman-VII solution (but the density is non-zero at the boundary) where
\begin{equation}
	a^2=\frac{3}{4\pi\rho_0},\quad b^2=\frac{5r_b^2}{4\pi(\rho_0-\rho_b)}.
\end{equation}
From the field equations
\begin{eqnarray}
	p&=&\left(\frac{8\gamma^2}{b^2}\right)r^4+\left({2 \over b^2}-{8\gamma^2\over a^2}\right)r^2+\left(4\gamma^2-\frac{2}{a^2}\right) \label{radial_pressure}\\
	P&=&\left(\frac{8\gamma^4}{b^2}\right)r^6+\left({16\gamma^2 \over b^2}-{8\gamma^4\over a^2}\right)r^4+\left({4\over b^2}+4\gamma^4-{12\gamma^2 \over a^2}\right)r^2+\left(4\gamma^2-\frac{2}{a^2}\right). \label{tangential_pressure}
\end{eqnarray}
We take $\gamma=1$ for simplicity.
Thus, $p(0)=P(0)>0$ if
\begin{equation}
	a^2>\frac{1}{2}\quad \text{or} \quad \rho_0<\frac{3}{2\pi}.
\end{equation}
Since the radial pressure is of degree four in $r$ and is symmetric under the transformation $r\to -r$, equation $p(r)=0$ has a maximum of two roots for $r\geq 0$. If $r_b$ is the smaller root, $\rho_b$ can be found in terms of $\rho_0$ and $r_b$.
One way to find specific solution would be to choose $\rho_0$ first, and then choose $r_b$ such that $0\leq \rho_b<\rho_0$.
Here, we directly choose values of $a^2$ and $b^2$ such that all physical conditions, except (\ref{sound_speed}), are satisfied.

\begin{figure}[h!]
  \makebox[\textwidth][c]{\includegraphics[width=1.15\textwidth]{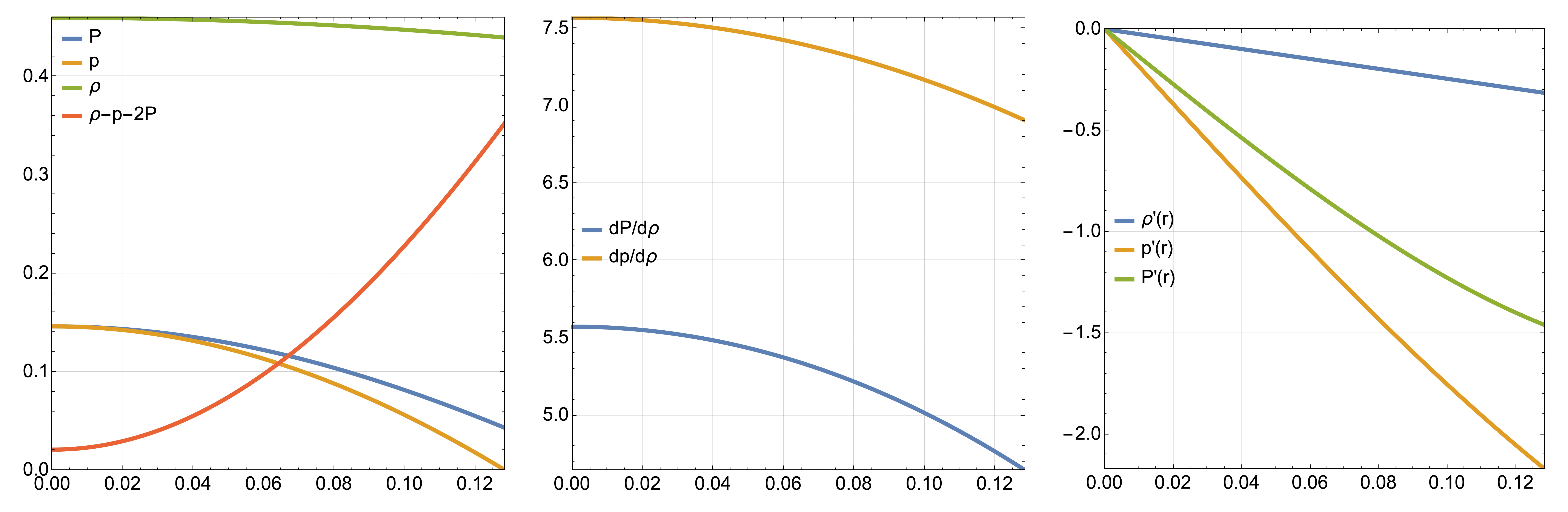}}
  \caption{
  Here, $\gamma^2=1$, $a^2=0.519$ and $b^2=0.325$. The boundary is at $r_b=0.129$ and the total mass is $M=0.004$. Therefore, the boundary in terms of total mass is given by $r_b=(32.269)M$. All the physical conditions, except (\ref{sound_speed}), are satisfied.}
\end{figure}

\section{Conclusion}\label{sec:vii}

We studied all possible combinations of the basic functions in curvature coordinates to produce all solutions of the (an)isotropic fluid system via quadrature and paid particular attention to sub-algorithms that can generate all (and only) regular solutions. This first required us to revisit and take a closer look at the geometric conditions of central regularity under the Einstein equations of the system and work out different equivalent conditions of regularity.  The conditions of regularity, in general, are three conditions on two functions, with the only exception in the case of isotropy when two conditions on the potential function are found to be sufficient. We also found that one can allow $p(r)$ to be non-differentiable at $r=0 $ as long as $\lim_{r\to 0} r p'(r)=0$  to generate regular solution; one should include such functions to generate all (regular) solutions when using $p(r)$ as one of the inputs.

There is no discussion in the literature on the regularity of the existing algorithm for the anisotropic system \cite{herrera2008all}. We, therefore, revisited this first using a slightly different set of variables, $\Phi'(r)$ and $P(r)$. We noted that the parameter arising in the general solution of this algorithm could be interpreted as the core mass in the presence of a nonzero core, which, therefore, cannot be negative or unbounded on physical grounds. For solutions starting from $r=0$, i.e., complete spheres (possibly terminating at a finite radius), the same analysis gives multiple solutions with zero central mass, and they are regular if and only if the parameter is set to zero. Thus, all anisotropic spheres are produced in a 1-1 fashion using $\Phi'(r)$ and $P(r)$ as generating functions. When $\Phi'(r)$ and $\chi(r)$ are used as input functions, as in \cite{herrera2008all}, the parameter in the general solution reverses its role and does not have to be zero; one, thus, have a one-parameter family of solutions that are regular under the usual initial conditions on $\Phi'(r)$ and $\chi(r)$ for regularity. In addition, in comparison with the algorithm using $\Phi'(r)$ and $P(r)$, we can now interpret the parameter as the central density. This also means that for pure isotropy, we get a one-parameter family of regular solutions, and the parameter, the central density, remains free; this analysis thus complements the analysis for the isotropic system in \cite{PhysRevD.67.104015}.

We found that the three other combinations---$(\rho(r)$, $p(r)$), $(\rho(r),\Phi'(r))$, and $(p(r), \Phi'(r))$---lead to another algorithm for the anisotropic system. The equivalency of the three pairs in this algorithm arises precisely because of the same algebraic relation between these three functions that made the different equivalent sets of regularity conditions possible.\footnote{We could not find a formal statement in the literature to this effect, but that $(\rho(r), p(r))$ can serve as generating functions for all anisotropic solutions is amply clear if one reads the generalized TOV equation as the defining algebraic equation for $P(r)$ in terms of $\rho(r)$ and $p(r)$.This fact, for example, was noted in \cite{boonserm2007solution} and this was used in \cite{lake2009generating} to generate a certain class of anisotropic solutions (using the Newtonian relationship $p'=-m\rho/r^2$).}
The line element (\ref{line_elementsupersimplified}) is very suitable for this algorithm. We noted that in regular solutions finiteness of $p(0)$ has to be be supplemented by $\lim_{r\to 0} r p'(r)=0$. This extra condition comes from our analysis of (\ref{momentum_conservation}) and is reconfirmed by the tensorial conditions of regularity. As for the other two choices---$(\rho(r), \Phi'(r))$ and $(p(r), \Phi'(r))$---we find that they are algebraically equivalent to $(\rho(r)$, $p(r)$) and can be used with their accompanying regularity conditions to generate all regular solutions. It is quite conceivable that one would convert any such pair to $(\rho(r)$, $p(r)$) and use (\ref{line_elementsupersimplified}). Note that $\rho(r)$ and $p(r)$ are output variables in the existing algorithms of isotropic and anisotropic systems and as such the line element (\ref{line_elementsupersimplified}) is especially significant only in this new algorithm for the anisotropic system. The conditions of regularity, as well as any energy condition, become a simple matter of choosing the appropriate algebraic forms of two generating functions in this algorithm. These exhaust all possible combinations of the basic functions of the anisotropic system. These results can be used with equations of state with little change.

An interesting feature of a Riccati equation is that if a particular solution is known one can work out the general solution. We have now three algorithms---with $(\rho(r)$, $p(r)$), $(\Phi'(r), P(r))$, and $(\Phi'(r), \chi(r))$---that generate all solutions and each of them is capable of generating all regular solutions. Any given solution generated by them can be seen as the particular solution of the Riccati equation (in $p(r)$ or in $\Phi'(r)$) and one gets two distinct one-parameter family of solutions. We found that one of them necessarily results in solutions that are regular at the center, for both isotropy and anisotropy, if the particular solution used is regular.

Finally, following the new algorithm, we used $(\rho(r), \Phi(r))$ and $(\rho(r), p(r))$ to generate examples of regular solutions that satisfy other physical conditions in addition to the condition of central regularity. In another work, we will discuss various maps between and within isotropic and anisotropic solutions and see how they can facilitate the process of finding solutions that satisfy all physical conditions.

\section*{Acknowledgements}
We thank Pankaj Joshi, Kayll Lake, and Malcolm MacCallum for useful communications. Special thanks to Charlie Brewer for useful comments on the manuscript. RS acknowledges the support of a Julia Williams Van Ness Merit Scholarship.

\end{document}